\documentclass{mem}
\usepackage{natbib}
\usepackage{txfonts}
\usepackage{balance}
\usepackage{graphicx}
\usepackage[a4paper]{hyperref}
\idline{1}{1}
\begin{document}

\title{
Toward the analysis of waves in the solar atmosphere based on NLTE spectral synthesis from 3D MHD simulations
}

   \subtitle{}

\author{
M. \, Haberreiter\inst{1} 
\and W. \, Finsterle\inst{2}
\and S. \, McIntosh\inst{3}
\and S. \, Wedemeyer-B{\"o}hm\inst{4,5}
}

  \offprints{M. Haberreiter}

\institute{
Laboratory for Atmospheric and Space Physics, University of Colorado,
1234 Innovation Drive, Boulder, CO, 80303, USA, \email{haberreiter@lasp.colorado.edu}
\and
Physikalisch-Meteorologisches Observatorium Davos, World Radiation Center,
Dorfstrasse 33, CH-7260 Davos Dorf, Switzerland
\and
High Altitude Observatory, National Center for Atmospheric Research,
P.O. Box 3000, Boulder, CO, 80307
\and
Institute of Theoretical Astrophysics, University of Oslo, P.O. Box 1029 Blindern, N-0315 Oslo, Norway
\and
Center of Mathematics for Applications CMA, University of Oslo, Box 1053 Blindern,
N-0316 Oslo, Norway
}

\authorrunning{Haberreiter et al.}
\titlerunning{Toward NLTE spectral synthesis from 3D MHD simulations}

\abstract{
From the analysis of Dopplergrams in the K~{\sc{i}}~7699~{\AA} and Na~{\sc{i}}~5890~{\AA}  spectral lines observed with the Magneto-Optical filter at Two Heights (MOTH) experiment during the austral summer in 2002--03 we find upward traveling waves in magnetic regions. Our analysis shows that the dispersion relation of these waves strongly depends on whether the wave is detected in the low-beta or high-beta regime. Moreover, the observed dispersion relation does not show the expected decrease of the acoustic cut-off frequency for the field guided slow magnetic wave. Instead, we detected an increase of the travel times below the acoustic cut-off frequency and at the same time a decrease of the travel time above it. To study the formation height of the spectral lines employed by MOTH in greater detail we are currently in the process of employing 3D MHD simulations carried out with CO$^5$BOLD to perform NLTE spectral synthesis. 
\keywords{Sun: chromosphere -- Sun: faculae, plages -- Sun: oscillations -- Sun: activity -- Line: profiles -- Magnetic fields -- Magnetohydrodynamics (MHD) -- Waves 
}
}
\maketitle{}

\section{Introduction}
The physics of the chromosphere and corona is most important for an understanding of the formation and variability of the solar UV/EUV radiation. However, the heating mechanism that causes the rising temperature profile of the upper chromosphere is not fully understood. In particular, the reason why the Sun shows a rising temperature profile over all phases of the solar cycle is still an open question. 

A number of potential heating mecha\-nisms of the solar chromos\-phere and co\-ro\-na are e.g. low-fre\-quency magne\-to\-acous\-tic gra\-vity (MAG) waves \citep{MH-Jefferies2006ApJL,MH-McIntosh2006ApJL}, the Farley-Bune\-man insta\-bility \citep{MH-Fontenla2008}, co\-ronal heating from Alfv\'en waves \cite{MH-Cranmer2007}, nano\-flares \cite{MH-PatsourakosKlimchuk2009}, and heating processes of the co\-rona are rooted in the chromos\-phere \citep{MH-DePontieu2009}.

\begin{figure}
\begin{center}
\resizebox{0.5\textwidth}{!}{\includegraphics[clip=true]{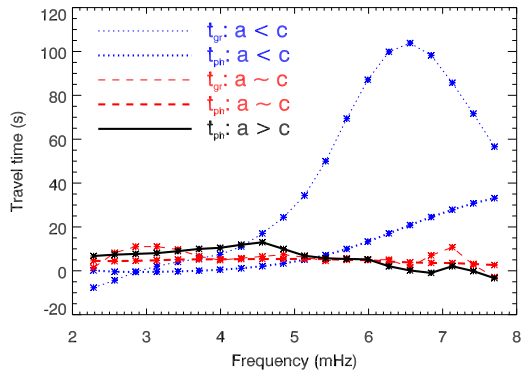}}
\caption{\footnotesize
Observed dispersion relations from the analysis of the Doppler signal observed with MOTH. The thin lines denote the phase travel time and the thick lines the group travel times. The different linestyles denote whether the travel time was derived from a magnetic or non-magnetic region, i.e. the dotted line corresponds to areas on the solar disk where $a<c$, the dashed lines where $a\sim c$, and the solid line where $a>c$. For details see \cite{MH-HaberreiterFinsterle2010}.}
\label{Haberreiter-fig:DISPREL}
\end{center}
\end{figure}

The ultimate aim of this work is to study the energy provided by MAG waves. Theory predicts that different types of MAG waves exist in the solar atmosphere \citep{MH-Schunker2006,MH-Cally2007}. They can couple at the height where the magnetic pressure equals the gas pressure, exchange energy, and thereby convert into a different wave type. Moreover, due to partial mode conversion, theory predicts that more than one wave type is likely to exist on both sides of the conversion layer. These processes are already well understood from a theoretical point of view, however mode conversion has not yet been unambiguously identified in observations.

The unambiguous identification of the different wave types from the observations is however a challenging task. To clearly disentangle the various wave types it is crucial to determine the observational heights of the instrument with respect to the magnetic canopy, as has been pointed out by \cite{MH-Bogdan2003}. From the Magneto-Optical Filter at Two Heights experiment \citep[MOTH]{MH-Finsterle2004SoPh} and the Michelson Doppler Imager observations \citep[MDI]{MH-Scherrer1998} onboard the NASA/ESA Solar and Heliospheric Observatory (SOHO), \cite{MH-HaberreiterFinsterle2010} determined the group and phase travel times of the observed MAG waves depending on the $a=c$-layer, i.e. the layer where the sound speed equals the Alfv\`en speed, with respect to the observational height of the K\,{\sc{i}} 7699\,{\AA} and Na\,{\sc{i}} 5890\,{\AA} lines. Fig.\,\ref{Haberreiter-fig:DISPREL} gives the dispersion relation for the cases where $a>c$, $a \sim c$, and $a<c$; for details see \cite{MH-HaberreiterFinsterle2010}.

From this analysis it is clear that the conversion layer changes the type of wave that is observed. The main question that needs to be answered is to what extent is the wave converted from a slow acoustic mode to a fast magnetic mode \citep{MH-Cally2007}. The observed dispersion relation has the potential to shed light on this matter. However, the dispersion relations derived from the observations can only be fully understood through the analysis of wave propagation from numerical simulations. We will use 3-dimensional (3D) MHD simulations carried out with CO$^5$BOLD \citep{MH-Freytag2002,MH-Wedemeyer2004,MH-Schaffenberger2006,MH-Steiner2007AN}. 

In this paper we report on continuing efforts to provide synthetic data that might allow us to understand the observed dispersion relations of these waves. 
\begin{figure*}
\resizebox{0.5\textwidth}{!}{\includegraphics[clip=true]{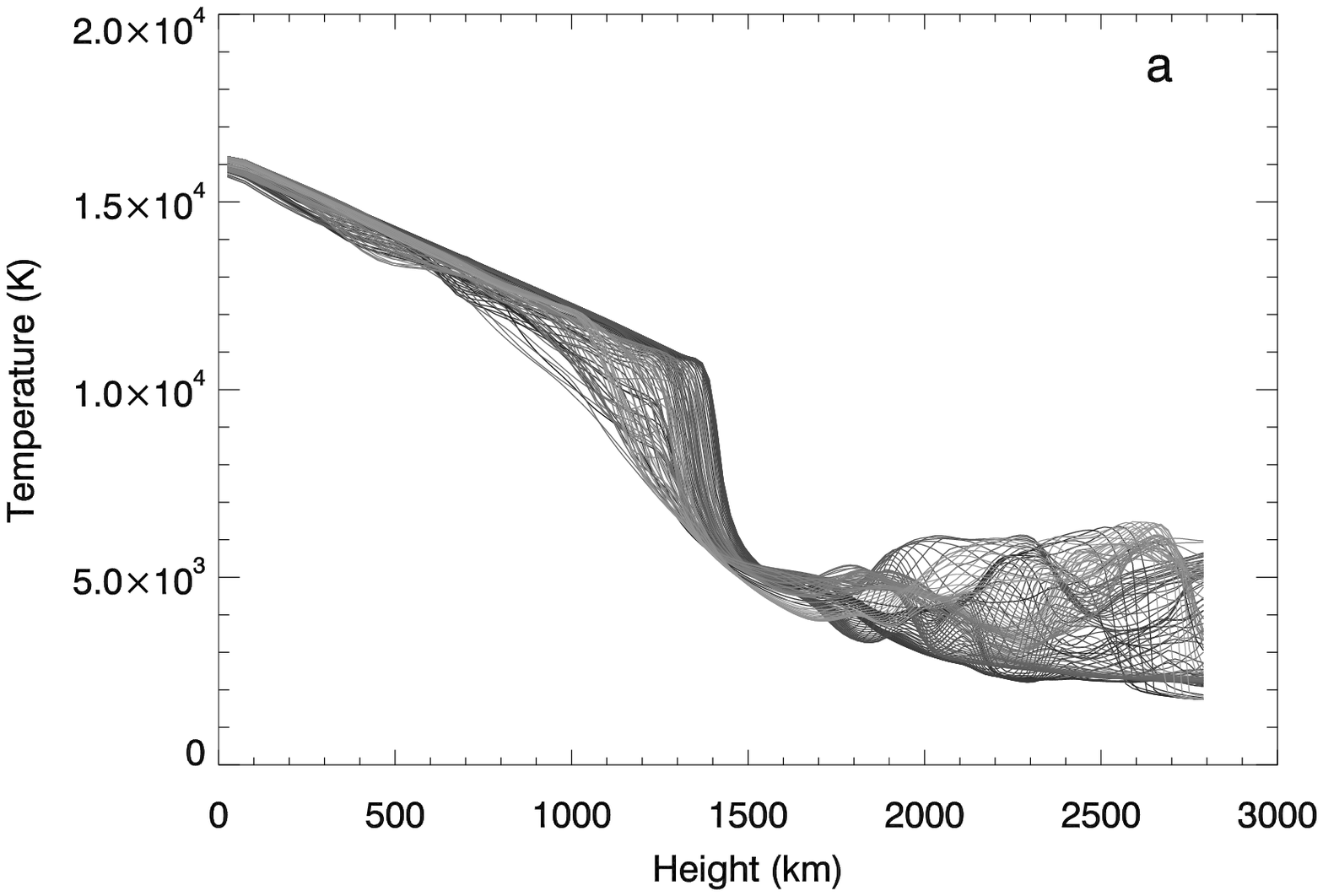}}
\resizebox{0.495\textwidth}{!}{\includegraphics[clip=true]{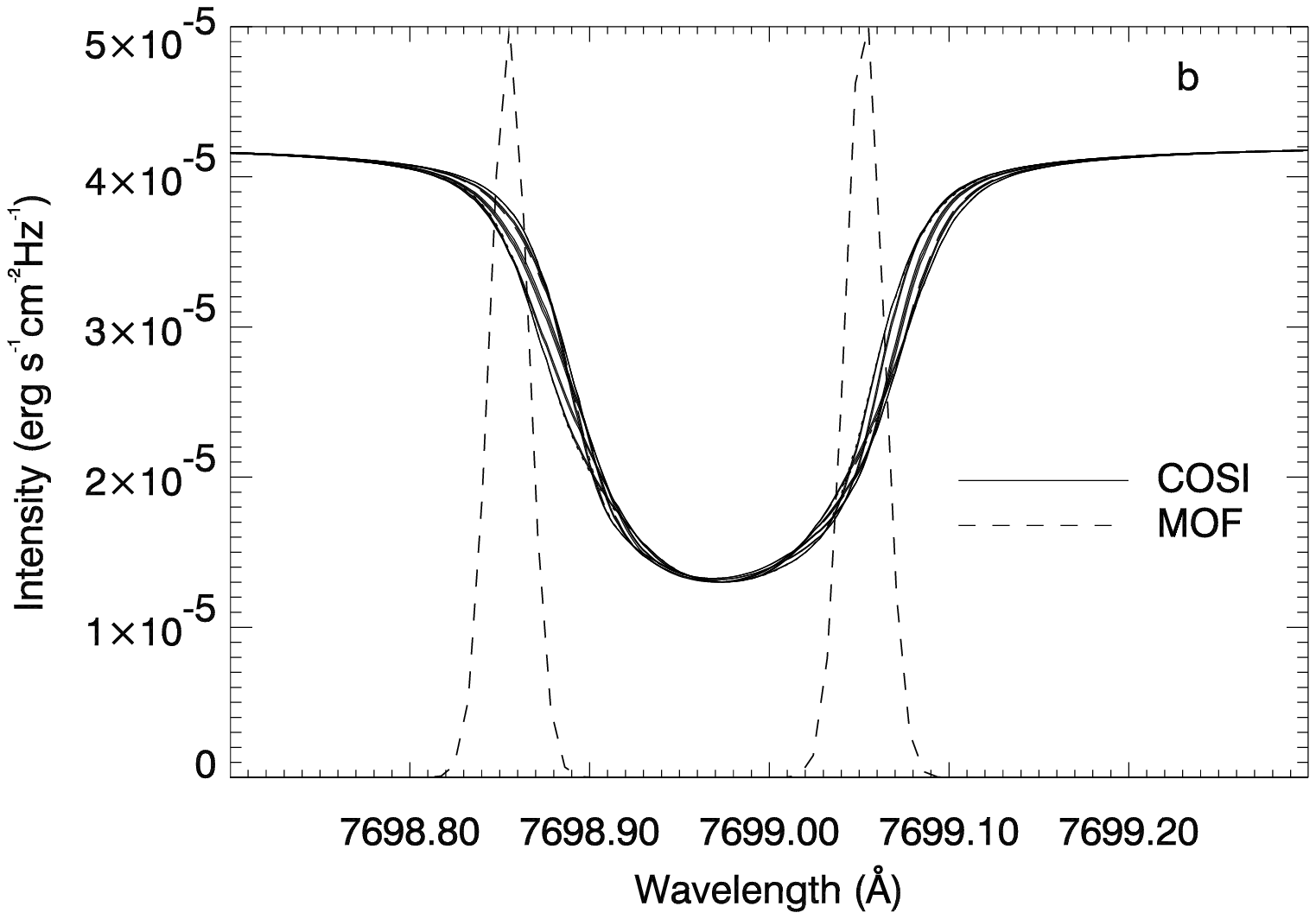}}
\caption{\footnotesize
{\em Panel (a):} Shown are the temperature structures of a snapshot of the simulation carried out with CO$^5$BOLD. {\em Panel (b):} Shown is the modulation of the K~{\sc{i}}~7699~{\AA} line calculated with COSI based on the Doppler signal of a vertically propagating wave with $\nu$=14\,mHz in a 1D atmosphere structure. The dashed line indicates the narrow band MOF filters employed with the MOTH instrument; for details see \cite{MH-Haberreiter2007AN}.}
\label{Haberreiter-fig:TEMP}
\end{figure*}
\section{Ongoing spectral synthesis from 3D MHD simulations}

Fig.\,\ref{Haberreiter-fig:TEMP} shows the temperature profiles of a snapshot of a vertical slice through the simulation carried out with CO$^5$BOLD. We are currently in the process of calculating the NLTE level populations of hydrogen, sodium, and potassium, and the most important elements to synthesize the solar line profiles for each of the grid points and snapshots. This will allow us to study in detail the formation height of these lines, which is essential for an understanding of the phase lag between the different observational layers in the solar atmosphere. 
 
\section{Results from 1D model}

For testing reasons we have already modeled the K~{\sc{i}}~7699~{\AA} line profile being modulated with the Doppler velocity signal corresponding to waves with various frequencies. The line profiles have been calculated with the COde for Solar Irradiance \cite[COSI]{MH-Haberreiter2008b}. Fig.\,\ref{Haberreiter-fig:TEMP}b shows the line profiles for waves with 14\,mHz. It is clear that the spectral line does not show a pure Doppler shift, but rather a change in the overall line profile, i.e. the C-shape of the line. The reason is that the core of the line is formed higher in the solar atmosphere than the wings of the line. Therefore, the line center and the wings of the line are not modulated simultaneously by the Doppler signal but experience a time lag. If the spatial wavelength of the acoustic wave is of the order of the extent of the contribution function of the spectral line, then the line profile is modulated, as shown in Fig.\,\ref{Haberreiter-fig:TEMP}b. This poses a challenge for the exact determination of the Doppler signal of high frequency waves from this kind of observations; for details see \cite{MH-Haberreiter2007AN}.

Detailed modeling of the wave excitation as carried out by \cite{MH-Steiner2007AN} and \cite{MH-Nutto2010} will allow us to disentangle the effect of the wavelength of the acoustic wave on the line profile.  

\section{Synergy of 1D and 3D models}
Semi-empirical 1D models are very successful for modeling the solar spectrum from the UV/EUV to the IR \citep{MH-Fontenla2009b, MH-Haberreiter2008b,MH-Shapiro2009}. Of course, 1D models do not represent any dynamics of the solar atmosphere, but are a temporal and spatial mean of the solar atmosphere. However, as they can reproduce a broad wavelength range and also detailed line profiles, they are a powerful means to study realistic radiative losses of different layers in the solar atmosphere, in particular the chromosphere, for the quiet Sun as well as active regions.

In the near future we plan to synthesize detailed line profiles as well as the spectrum over broad wavelength ranges based on 3D MHD simulations. Hopefully, 1D and 3D models can provide complementary insight to achieve this goal. For example, detailed radiative losses calculated from 1D models might prove valuable for a comparison with the values used in 3D MHD simulations.

\section{Discussion and Conclusions}
From the analysis of Dopplergrams in the K~{\sc{i}}~7699~{\AA} and Na~{\sc{i}}~5890~{\AA} spectral lines observed with the MOTH experiment during the austral summer in 2002--03 we find upward traveling waves in magnetic regions. Our analysis shows that the dispersion relation of these waves strongly depends on whether the wave is detected in the low-$\beta$ or high-$\beta$ regime. Moreover, the observed dispersion relation does not show the expected decrease of the acoustic cut-off frequency for the field guided slow magnetic wave. Instead, we detect an increase of the travel times below the acoustic cut-off frequency and at the same time a decrease of the travel time above it. To study the formation height of the spectral lines employed by MOTH in greater detail, we use 3D MHD simulations carried out with CO$^5$BOLD to carry out NLTE spectral synthesis. 

Total and spectral solar irradiance changes are understood to be linked to the Sun's magnetic cycle; see e.g.\,\cite{MH-Haberreiter2010IAU}. With increasing magnetic activity more radiation is emitted, in particular in the UV and EUV. However, the physical processes that provide the enhanced heating in magnetic regions are not yet fully understood. Waves {\em leaking} through magnetic regions are considered one of the drivers. A detailed spectral synthesis based on different scenarios of magnetic field strengths representing different phases during the solar cycle might help to understand the contribution of MAG waves in the heating process. 
{\small
\begin{acknowledgements}
{MDI data courtesy of the SOHO/MDI Consortium. SOHO is a project of international cooperation between ESA and NASA. The MOTH experiment was funded by award OPP-0087541 from the National Science Foundation (NSF).}
\end{acknowledgements}}


\begin{thebibliography}{20}
\expandafter\ifx\csname natexlab\endcsname\relax\def\natexlab#1{#1}\fi

\bibitem[{{Bogdan} {et~al.}(2003){Bogdan}, {Carlsson}, {Hansteen}, {McMurry},
  {Rosenthal}, {Johnson}, {Petty-Powell}, {Zita}, {Stein}, {McIntosh}, \&
  {Nordlund}}]{MH-Bogdan2003}
{Bogdan}, T.~J., {Carlsson}, M., {Hansteen}, V.~H., {et~al.} 2003, \apj, 599,
  626

\bibitem[{{Cally}(2007)}]{MH-Cally2007}
{Cally}, P.~S. 2007, Astronomische Nachrichten, 328, 286

\bibitem[{{Cranmer} {et~al.}(2007){Cranmer}, {van Ballegooijen}, \&
  {Edgar}}]{MH-Cranmer2007}
{Cranmer}, S.~R., {van Ballegooijen}, A.~A., \& {Edgar}, R.~J. 2007, \apjs,
  171, 520

\bibitem[{{De Pontieu} {et~al.}(2009){De Pontieu}, {McIntosh}, {Hansteen}, \&
  {Schrijver}}]{MH-DePontieu2009}
{De Pontieu}, B., {McIntosh}, S.~W., {Hansteen}, V.~H., \& {Schrijver}, C.~J.
  2009, \apjl, 701, L1

\bibitem[{{Finsterle} {et~al.}(2004){Finsterle}, {Jefferies}, {Cacciani},
  {Rapex}, {Giebink}, {Knox}, \& {Dimartino}}]{MH-Finsterle2004SoPh}
{Finsterle}, W., {Jefferies}, S.~M., {Cacciani}, A., {et~al.} 2004, \solphys,
  220, 317

\bibitem[{{Fontenla} {et~al.}(2009){Fontenla}, {Curdt}, {Haberreiter}, \&
  {Harder}}]{MH-Fontenla2009b}
{Fontenla}, J.~M., {Curdt}, W., {Haberreiter}, M., \& {Harder}, J. 2009, \apj,
  707, 482

\bibitem[{{Fontenla} {et~al.}(2008){Fontenla}, {Peterson}, \&
  {Harder}}]{MH-Fontenla2008}
{Fontenla}, J.~M., {Peterson}, W.~K., \& {Harder}, J. 2008, \aap, 480, 839

\bibitem[{{Freytag} {et~al.}(2002){Freytag}, {Steffen}, \&
  {Dorch}}]{MH-Freytag2002}
{Freytag}, B., {Steffen}, M., \& {Dorch}, B. 2002, Astronomische Nachrichten,
  323, 213

\bibitem[{{Haberreiter}(2010)}]{MH-Haberreiter2010IAU}
{Haberreiter}, M. 2010, in Proceedings IAU Symposium No. 264, 2009. Edited by
  A.G. Kosovichev, A.H. Andrei, and J.-P. Rozelot, Cambridge: Cambridge Univ.
  Press, 231

\bibitem[{{Haberreiter} \& {Finsterle}(2010)}]{MH-HaberreiterFinsterle2010}
{Haberreiter}, M. \& {Finsterle}, W. 2010, \solphys, accepted

\bibitem[{{Haberreiter} {et~al.}(2007){Haberreiter}, {Finsterle}, \&
  {Jefferies}}]{MH-Haberreiter2007AN}
{Haberreiter}, M., {Finsterle}, W., \& {Jefferies}, S.~M. 2007, Astronomische
  Nachrichten, 328, 211

\bibitem[{{Haberreiter} {et~al.}(2008){Haberreiter}, {Schmutz}, \&
  {Hubeny}}]{MH-Haberreiter2008b}
{Haberreiter}, M., {Schmutz}, W., \& {Hubeny}, I. 2008, \aap, 492, 833

\bibitem[{{Jefferies} {et~al.}(2006){Jefferies}, {McIntosh}, {Armstrong},
  {Bogdan}, {Cacciani}, \& {Fleck}}]{MH-Jefferies2006ApJL}
{Jefferies}, S.~M., {McIntosh}, S.~W., {Armstrong}, J.~D., {et~al.} 2006,
  \apjl, 648, L151

\bibitem[{{McIntosh} \& {Jefferies}(2006)}]{MH-McIntosh2006ApJL}
{McIntosh}, S.~W. \& {Jefferies}, S.~M. 2006, \apjl, 647, L77

\bibitem[{{Nutto} {et~al.}(in press, see this volume) {Nutto}, {Steiner}, \& {Roth}}]{MH-Nutto2010}
{Nutto}, C., {Steiner}, O., \& {Roth}, M., in press, Mem. S. A. It. [this volume]


\bibitem[{{Patsourakos} \& {Klimchuk}(2009)}]{MH-PatsourakosKlimchuk2009}
{Patsourakos}, S. \& {Klimchuk}, J.~A. 2009, \apj, 696, 760

\bibitem[{{Schaffenberger} {et~al.}(2006){Schaffenberger},
  {Wedemeyer-B{\"o}hm}, {Steiner}, \& {Freytag}}]{MH-Schaffenberger2006}
{Schaffenberger}, W., {Wedemeyer-B{\"o}hm}, S., {Steiner}, O., \& {Freytag}, B.
  2006, in Astronomical Society of the Pacific Conference Series, Vol. 354,
  Solar MHD Theory and Observations: A High Spatial Resolution Perspective, ed.
  J.~{Leibacher}, R.~F. {Stein}, \& H.~{Uitenbroek}, 345

\bibitem[{{Scherrer}(1998)}]{MH-Scherrer1998}
{Scherrer}, P.~H. 1998, in ESA Special Publication, Vol. 418, Structure and
  Dynamics of the Interior of the Sun and Sun-like Stars, ed. S.~{Korzennik},
  15

\bibitem[{{Schunker} \& {Cally}(2006)}]{MH-Schunker2006}
{Schunker}, H. \& {Cally}, P.~S. 2006, \mnras, 372, 551

\bibitem[{Shapiro {et~al.}(2010)Shapiro, Schmutz, Sch\"oll, Haberreiter, \&
  Rozanov}]{MH-Shapiro2009}
Shapiro, A.~I., Schmutz, W., Sch\"oll, M., Haberreiter, M., \& Rozanov, E., \aap, submitted

\bibitem[{{Steiner} {et~al.}(2007){Steiner}, {Vigeesh}, {Krieger},
  {Wedemeyer-B{\"o}hm}, {Schaffenberger}, \& {Freytag}}]{MH-Steiner2007AN}
{Steiner}, O., {Vigeesh}, G., {Krieger}, L., {et~al.} 2007, Astronomische
  Nachrichten, 328, 323

\bibitem[{{Wedemeyer} {et~al.}(2004){Wedemeyer}, {Freytag}, {Steffen},
  {Ludwig}, \& {Holweger}}]{MH-Wedemeyer2004}
{Wedemeyer}, S., {Freytag}, B., {Steffen}, M., {Ludwig}, H., \& {Holweger}, H.
  2004, \aap, 414, 1121

\end{thebibliography}
\end{document}